\documentclass[10pt]{article}
\usepackage{amsmath,amssymb,amsbsy,amsthm}
\let\emptyset\varnothing
\let\epsilon\varepsilon
\let\phi\varphi

\def\st{\operatorname{St}}
\def\enc{\operatorname{StEnc}}
\def\dec{\operatorname{StDec}}

\def\N{\mathbb N}

\def\E{{\operatorname{\bf{E}}}}

\newtheorem{theorem}{Theorem}

\newtheorem{definition}{Definition}

\begin{document}

\title{Using Kolmogorov Complexity for Understanding Some Limitations on Steganography}

\author{Boris Ryabko\footnote{Institute of Computational Technologies of
Siberian Branch of Russian Academy of Science
Siberian State University of Telecommunications  and Informatics,  Novosibirsk, Russia
   boris@ryabko.net}, Daniil Ryabko
\footnote{INRIA Lille-Nord Europe
   daniil@ryabko.net}
    }

\maketitle

\begin{abstract}
Recently perfectly secure steganographic systems have been
described for a wide class of sources of covertexts. The speed of
transmission of secret information for these stegosystems is
proportional to the length of the covertext. In this work we show
that there are sources of covertexts for which such stegosystems
do not exist. The key observation is that if the set  of possible
covertexts has a maximal Kolmogorov complexity, then a high-speed
perfect stegosystem has to have complexity of the same order.
\end{abstract}

\section{Introduction}

The goal of steganography can be described as follows. Alice and
Bob can exchange messages  of a certain kind (called covertexts)
over a public channel. The covertexts can be, for example, a
sequence of photographic images, videos, text emails and so on.
Alice wants to pass some secret information to Bob so that  Eve,
the observer, cannot notice that any hidden information is being
passed. Thus,  Alice should use the covertexts to hide the secret
text.

Cachin \cite{ca} suggested an information-theoretic model for
steganography, along with a definition of a perfectly secure
steganographic system. According to this model, Alice has an
access to a probabilistic source of covertexts.  It is usually
assumed that the secret message can be represented as a sequence
of independent equiprobable  bits.  She has to embed her secret
message into covertexts in such a way that Bob can decode the
message. A stegosystem is called perfectly secure if the
distribution of the output  is the same as the distribution at the
source of covertexts. Indeed, in this case nobody is able to
distinguish containers with hidden information  and "empty" ones
(i.e. without hidden information).

Let us consider an example of a perfect stegosystem suggested in
\cite{rr}. The source of covertexts $\mu$ is as follows. It
generates sequences of $n$ independently identically distributed
letters from some finite alphabet $A$, where $n\in\N$ is given.
For the sake of simplicity we consider the binary alphabets
$A=\{a,b\}$, but the  construction can be used for the general
case too \cite{rr}. The distribution $\mu$ can be unknown to Alice
and Bob.  Suppose that Alice has to transmit a secret sequence
$y^*=y_1y_2\dots$ generated by a source $\omega$, where
$\omega(y_i=0)=1/2$ independently for all $i\in\N$, and let there
be given a covertext sequence $x^*=x_1x_2\dots$ generated by
$\mu$. For example, let
\begin{equation}\label{eq:ex}
y^*=01100\dots, \ \ x^*=aababaaaabbaaaaabb 
\end{equation}
The sequences $x^*$ and $y^*$ are encoded in a new sequence $X$
(to be transmitted to Bob) such that $y^*$ is uniquely determined
by $X$, and the distribution of $X$ is the same as the
distribution of $x^*$ (that is, $\mu$; in other words, $X$ and
$x^*$ are statistically indistinguishable).

The encoding is carried out in two steps. First we group all
symbols of $x^*$ into pairs, and denote
$$
 aa=u,\ bb=u,\ ab=v_0,\ ba=v_1.
$$
In our example, the sequence~(\ref{eq:ex}) is represented as
$$
x^*=aa\,ba\,ba\,aa\,ab\,ba\,aa\,aa\,bb
=uv_1v_1uv_0v_1uuu
$$
Then $X$ is acquired from $x^*$ as follows: all pairs
corresponding to $u$ are left unchanged, while all pairs
corresponding to $v_k$ are transformed to pairs corresponding to
$v_{y_1}v_{y_2}v_{y_3}$; in our example
$$
 X= aa \, ab \, ba \, aa \, ba \, ab \,  aa \,  aa \,  bb
$$
Decoding is obvious: Bob groups the symbols of $X$ into pairs,
ignores all occurrences of $aa$ and $bb$ and changes $ab$ to $0$
and $ba$ to $1$.

The described stegosystem has the following properties.
 The sequence of symbols output by the stegosystem obeys the same
distribution $\mu$ as the input sequence. The average length of
the transmitted secret sequence is $n\mu(ab)$; in other words, the
speed of transmission of hidden information is $\mu(ab)$ secret
bits per letter of covertext. Moreover, in \cite{rr} a generalization of the
described construction is proposed, for which the speed of
transmission of secret text approaches
the Shannon entropy
$h(\mu)=-(\mu(a)\log \mu(a)+\mu(b)\log \mu(b))$ when $n$ goes to
infinity. In addition, a similar construction is proposed in the
same work for the case of arbitrary alphabets and for finite-alphabet  Markov sources of covertexts. 

So, we can see that perfectly secure  stegosystems exist for a
wide class of covertexts and, moreover, such systems are quite
simple and have a high speed of transmission of secret
information. Naturally, one is interested in the question of
whether such stegosystems exist for any possible 
source of covertext. This problem is of interest since sources of
covertexts that are of particular practical importance, such as
texts in natural languages or photographs, do not seem to be
well-described by any known simple model. Here we answer this
question in the negative. More precisely, it turns out that there
exists such a set of covertexts of length $n$ for which
 simple stegosystems which have speed of transmission of hidden text  $\Omega(n)$ do not exist. Here  simplicity
is measured by Kolmogorov complexity  of the system and a
stegosystem is ``simple'' means that its complexity is $\exp(
o(n))$, when $n$ goes to infinity. Kolmogorov complexity  is an
intuitive notion that often helps to establish results that help
to understand the principled limitations a certain problem or
model imposes; it has been used as such in many works, see, for
example, \cite{Vi,USS, V-L, VVV}.

This result can be interpreted as that there are such complicated sources
of data, that one cannot conceivably put significantly more information into a source,
without changing its characteristics, even though the entropy of the source is very high.
This may explain what is known in practice; for example, it is apparently very hard to put any hidden
message into a given text in a natural language, without making the text ``unnatural''. Of course, rather
than trying to change a given text,  the communicating
parties can easily agree in advance on two  texts each of which codes one secret bit, so that when the need
for communication arises, Alice can transmit one of the texts thereby passing one bit. However,
in order to communicate more than one bit, to use the same method they would have to have
a database of covertexts that is exponentially large with respect to the message to pass. Moreover,
even this stegosystem will not be perfectly secure, since the source of covertexts with hidden
information is concentrated on a small subset of all the possible covertexts of the given length.
If the stegosystem is used once, then perhaps no reliable detection of the hidden message is possible.
If it is to be used on multiple occasions, that is, if we wish to construct a general purpose stegosystem
for transmitting, say, $\delta n$ bits with an $n$-bit message (for some fixed $\delta$), we will need
to construct a database of effectively all possible covertexts. At least, this is the case for some sources
of covertexts, as the result of this work demonstrates, and it seems likely that it is the case for such
sources as texts in natural languages or even photographic images. Thus, our result may be helpfull
in clarifying the nature of the difficulties that arise in construction of real steganographic
systems which use human-generated sources of covertexts.

\section{Preliminaries}
We use the following model for steganography, mainly following
\cite{ca}. It is assumed that Alice  has an access to an oracle
which generates covertexts according to some fixed but unknown
{\em distribution of covertexts $\mu$}.
 Covertexts belong
to some finite alphabet $A$. For the sake of simplicity we
consider the case $A=\{0,1\}$; the general case is analogous.
Alice wants to use this source to transmit hidden messages. A {\em
hidden (or secret) message} is a sequence of letters from $\{0,1\}$
generated independently with equal probabilities of $0$ and $1$.
We denote the {\em source of hidden messages} by $\omega$. This is
a commonly used  model for the source of secret messages, since it
is assumed that secret messages are encrypted by Alice using a key
shared only with Bob. If Alice  uses the Vernam cipher  (a
one-time pad) then the encrypted messages are indeed generated
according to the Bernoulli $1/2$ distribution, whereas if Alice
uses modern block or stream ciphers then the encrypted sequence ``looks
like'' a sequence of random Bernoulli $1/2$ trials. (Here
 ``looks like'' means indistinguishable in polynomial time,
or that the likeness is confirmed experimentally by
 statistical data, see, e.g. \cite{rf}.)
The third party, Eve is a passive adversary: Eve  is reading all
messages passed from Alice to Bob and is trying to determine
whether secret messages are being passed in the covertexts or not.
Clearly,  if covertexts with and without hidden information have
the same probability distribution ($\mu$) then  it is impossible
to distinguish them.

Since the number of possible covertexts $x$ in the set $A^n$ is
finite, using a stegosystem once Alice can only transmit a finite
number of bits of  the  secret message.  We tacitly assume that
there are always more secret bits than Alice wants to pass, which
is formalized by assuming that Alice has an infinite secret
message (in practice, if Alice runs out of secrets, she can fill
the remainder of the message with random noise). Depending on the
covertext that Alice has and on the actual secret message, the
length of the secret text that she transmits may vary. Naturally,
one wishes to maximize the expected length of the secret message
that the encoder can transmit. We require, however, that the
decoding is always correct, that is, Bob gets the whole secret
message that Alice has transmitted, without errors.

The steganographic protocol can be summarized in the following
definitions.

\begin{definition}[secret or hidden text]
A source $\omega$ of secret  text $y^*=y_1,y_2,\dots$ is such that
$\omega (y_i=0)=\omega (y_i=1)=1/2$, independently for all
$i\in\N$.
\end{definition}

\begin{definition}[
stegosystem] A 
 stegosystem $\st$ is a family (indexed by $n$)
of pairs of functions: the encoder, that maps a pair $(x,y^*)\in
A^n\times \{0,1\}^\infty$ (a covertext and a secret sequence) into
a pair $(t,\enc_n(x,(y_1\dots y_t))\in\N\times A^n$: the number of
secret bits transmitted and the output covertext. The decoder
$\dec_n$ is a function from $A^n$ to $\{0,1\}^*$. We will often
omit the parameter $n$ from the notation, when its value is clear.
\end{definition}

\begin{definition}[
 steganographic protocol]
A parameter $n$ is fixed. Alice draws a   covertext $x\in A^n$
generated by a  source of covertexts $\mu$ (a distribution on
$A^n$) and a secret message $y^*=y_1,y_2,\dots$ according to the
source $\omega$. The sources $\omega$ and $\mu$ are independent of
each other.


Given  $x\in A^n$ and $y^*$  Alice  using a {\em stegosystem
$\st$} obtains the number of secret bits she can pass $t(x,y^*)\ge0$,
and a covertext $\enc(x,(y_1\dots y_t))\in A^n$ that is
transmitted over a public channel to Bob. (Only $\enc(x,(y_1\dots
y_t))$ is transmitted; the number $t$ is not.)

Bob (and any possible observer Eve) receives $x'\in A^n$ and
obtains using the decoder $\dec$ the  resulting message
$\dec_n(x')=y_1\dots y_t$.
\end{definition}

\begin{definition}[perfect security]
 A 
 steganogrpahic system is called perfectly secure if the sequence of covertexts $x^*$ and the steganographic sequence $X$ have
the same distribution: $Pr_{\mu\times\omega}(\enc=x')=\mu(x')$ for
any  $x'\in A^n$, where the first probability is taken with
respect to the distribution of covertexts $\mu$ and that of secret
text $\omega$.
\end{definition}

\begin{definition}[speed of transmission] For a  stegosystem $\st$
the speed of  transmission of secret text $v_n(\st)$ is defined as
$\E_{\mu\times\omega} t(x,y^*) / n$ (the expectation is  with
respect to $\mu$ and  $\omega$).
\end{definition}

Note that often (in particular, in \cite{ca}) more general steganographic protocols are considered,
allowing for non-perfect security, transmission with errors, several draws from the source of covertexts, etc.
We have decided to concentrate on the simple model presented since it is rich enough for perfectly secure stegosystems  to exist,
for a wide classes of sources of covertexts (e.g. all finite-memory sources, \cite{rr}).
Some possible extensions are discussed in Section~\ref{ext}.



For definitions, notation, and an introduction to Kolmogorov
complexity, see \cite{Vi}. Informally, the Kolmogorov complexity,
or algorithmic entropy, can be defined as follows \cite{Vi1}: 
$K(x)$ of a string $x$ is the length (number of bits) of a
shortest binary program (string) to compute $x$ on a fixed
reference universal computer (such as a particular universal
Turing machine). Intuitively, $K(x)$ represents the minimal amount
of information required to generate $x$ by any effective process.
The conditional Kolmogorov complexity $K(x|y)$ of $x$ relative to
$y$ is defined similarly as the length of a shortest program to
compute $x$, if $y$ is furnished as an auxiliary input to the
computation.

  We will use
some simple properties of $K$, such as $K(s)\le |s|+c$ for any
word $s$, whose proofs can be found in e.g. \cite{Vi}. Here it is
worth noting that $K(s)$ does not take into account the time
or extra
memory  it takes to compute $s$.

\section{Main results}

\begin{theorem}\label{th:main}
For every $\delta>0$ there is a family indexed by $n\in\N$ of
distributions $P_n$ on $A^n$ with $h(P_n)\ge n-1$, such that every
stegosystem  $\st_n$ whose Kolmogorov complexity  satisfies $\log
K(St_n)= o(n)$ and whose speed of transmission of hidden text
$v_n(\st_n)$ is not less than $\delta$, is not perfectly secure
from some $n$ on.
\end{theorem}


\begin{proof} We will construct a sequence of sets
  $X_n$ of words of length $n$ whose  Kolmogorov complexity is
the highest possible, namely $2^{\Omega(n)}$. For each $n\in\N$,
the distribution $P_n$ is uniform on $X_n$. We will then show
that, in order to have the speed of transmission $\delta>0$ a
perfectly secure stegosystem must be able to generate a large
portion of the set $X_n$, for each $n$. This will imply that the
complexity of such a stegosystem has to  be $2^{\Omega(n)}$. The
latter implication will be shown to follow from the fact that, in
order to transmit some information, a stegosystem must replace the
input with some output that could have been generated by the
source; this, for perfectly secure stegosystems, amounts to
knowing at least a large portion of $X_n$.

Fix $n\in\N$ and let  $X\subset A^{n}$ be any set such that
$|X|=2^{n-1}$ and
\begin{equation}\label{kx}
 K(X)= 2^{n} (1 +o(1)).
\end{equation}
 The existence of such a set can be shown
by a direct calculation of the number of all subsets with
$2^{n-1}$ elements; the maximal complexity is equal (up to a
constant) to the $\log$ of this value.

Assume that there is a perfectly secure stegosystem $St_n$ for the
family $P_n$, $n\in\N$, and let the speed of transmission of
hidden text be not less than $\delta$. Define the set $Z$ as the
set of those words which are used as codewords $Z:=\{x\in A^n:
\dec(x)\ne\Lambda\}$. Since the expected speed of transmission of
hidden text is lower bounded by $\delta$, we must have $|Z|\ge
\delta 2^{n-1}$ (indeed, since every word codes at most $n-1$
bits, the expected speed of transmission must satisfy
$(n-1)\frac{|Z|}{2^{n-1}}\ge\delta n$). Since $\st$ is perfectly
secure $Z\subset X$. Let us lower-bound the complexity
$K(Z|X\backslash Z)$ of the set $Z$ given $X\backslash Z$. Given
the description of $X\backslash Z$ and the description of $Z$
relative to $X\backslash Z$ one can reconstruct $X$. That is why
$K(Z|X\backslash Z)\ge K(X)- K(X\backslash Z)+O(1).$ The size of
$X\backslash Z$ is not greater than $2^{n-1}(1-\delta)$. Hence,
\begin{equation}\label{2}
K(Z|X\backslash Z) \ge K(X) - \max_{|U|\le2^{n-1}(1 - \delta)}
K(U) +O(1).
\end{equation}
 The latter maximal complexity  can be calculated as
follows:
$$
\max_{|U|\le2^{n-1}(1-\delta)} K(U) = \log     {2^n \choose
2^{n-1} (1-\delta)} + O(1).
$$
 Applying the Stirling approximation for factorial we
obtain
$$
\max_{|U|\le 2^{n-1}(1 - \delta)} K(U) \le  2^{n}(1-\gamma)
(1+o(1)),
$$
where $\gamma=1-h({1-\delta\over 2},{1+\delta\over 2})$. From this
equality, (\ref{kx}) and (\ref{2})  we get
$$
K(Z)\ge\gamma 2^{n} (1+o(1)).
$$

Furthermore, define $Z_0$ as the set of words that code those
secret messages that start with 0, and $Z_1$ those that start
with~1:
\begin{equation}\label{z}
 Z_i:=\{x\in A^n: \dec(x)=iu, u\in \{0,1\}^* \}, i\in\{0,1\}.
\end{equation}
 Clearly, $Z=Z_1\cup Z_0$. Hence, $K(Z|X\backslash Z)\le
K(Z_0|X\backslash Z)+K(Z_1|X\backslash Z) + O(1)$, so that
$K(Z_i|X\backslash Z)\ge K(Z|X\backslash Z)/2 + O(1)$ for some
$i\in\{0,1\}$. Let this $i$ be 1. Thus,
\begin{equation}\label{kz1}
 K(Z_1|X\backslash Z)\ge \gamma 2^{n-1} (1+o(1)).
\end{equation}
 We will next show how to obtain
$Z_1$ from $Z\backslash Z_1$ and  the stegosystem $\st$, thus
arriving at a contradiction with the assumption that $\log
K(\st)=o(n)$.

For a set $T\subset X$ define
$$
\phi(T):=\{\enc(x,1u): x\in T, u\in \{0,1\}^*\}.
$$
 Since $\st$ is perfectly
secure, $\phi(T)\subset X$ for every $T\subset X$.
 Let $T_0=X\backslash Z_1$, and $T_k=T_{k-1}\cup\phi(T_{k-1})$.
 Since $X$ is finite and each $T_{k-1}$ is a subset of $T_k$, there must be such $k_0\in\N$ that
 $T_k=T_{k_0}$ for all $k>k_0$.
 There are two possibilities: either $T_{k_0}= X$ or $X\backslash T_{k_0}\ne\emptyset$.
  Assume the latter, and define
$Z_1'=X\backslash T_{k_0}$. Then to obtain an element of $Z_1'$ as
an output of the  stegosystem $\st$, the input must be an element
of $Z_1'$ and a secret message that starts with $1$. From this,
and from the fact that the distribution of the output is the same
as the distribution of the input (that is, $\st$ is perfectly
secure), we get
$$
P_n(Z_1')=P_n(Z_1', y=1u)= P_n(Z_1')\omega(1)=P_n(Z_1')/2,
$$
which implies $P_n(Z_1')=0$ and $Z_1'=\emptyset$. Therefore, there
is a $k\in\N$ such that $T_k= X$. This means that a description of
$Z_1$ can be obtained from a description of $X\backslash Z_1=T_0$
and $\st$. Indeed, to obtain $Z_1$ it is sufficient to run $\enc$
on all elements of $T_0$ with all inputs starting with $1$, thus
obtaining $T_1$, and then repeat this procedure until we get
$T_{k+1}=T_k$ for some $k$, wherefrom we know that $T_k=X$ and
$Z_1=T_k\backslash T_0$. Thus,
\begin{equation}\label{contr}
 K(Z_1|X\backslash Z_1)\le
K(\st)+O(1)=2^{o(n)}
\end{equation}
 which contradicts~(\ref{kz1}).
\end{proof}

\section{Possible extensions}\label{ext}
The definitions of stegosystems and steganographic protocol that we have used  allow for several
extensions. In particular, we have made the  requirement
that Alice can draw only one covertext from the source, in order to construct
her message. We have also required that the decoding is always correct,
%
  did not allow for a  secret key in the
protocol (a secret key could be used before entering into
steganographic communication in order to obtain the secret message
$y^*$, but is out of scope of the protocol), etc.
These requirements, along with the requirement of perfect
security, might be considered restrictive; however, as was
mentioned in the Introduction, for some sources of covertexts
(such as i.i.d. or finite-memory sources) there are indeed
perfectly secure 
 steganographic systems that meet all the requirements we have made, and which also have the highest possible speed
of transmission of hidden text: $v_n(\st)$ approaches
(exponentially fast) the Shannon entropy $h(\mu)$ of the source of
covertexts, as $n$ grows (see \cite{rr}).
This  is why we have decided to sacrifice the generality for the sake of simplicity of the model presented.

Nevertheless, it is worth noting that  the main
results  of this work can be extended to more general cases.
For example, if we allow Alice and Bob to share a secret key $k_n$, then
trivially Theorem~1 holds with $K(\st_n)$ replaced by $K(\st_n)+K(k_n)$.
Let us briefly stop on another extension of the protocol.
Instead of allowing Alice to draw only one covertext from the source, we can allow her to draw several, say, $M$ covertexts.
Given $M$ covertexts $x_1,\dots,x_m$, where $x_i\in A^n$, and a secret sequence $y^*$ Alice constructs
a single $x'\in A^n$ which is passed (over  a public channel) to Bob. In particular, depending on the message $y^*$, Alice can chose
 $x_i$ that already encodes the message, if such $x_i$, $1\le i\le M$ exists. The speed $v_n$ of transmission
of secret text is measured with respect to what is passed over the public channel only (i.e. $x'$).
Then Theorem~\ref{th:main} admits the following extension: there are such sources of covertexts, that any perfectly secure simple stegosystem must draw $M_n=2^{\Omega(n)}$ covertexts in order to transmit $\delta n$ bits, for any given $\delta>0$.


\end{document}